\documentclass{aastex}
\usepackage{spr-astr-addons}
\usepackage{url}
\usepackage{amsfonts}
\usepackage{amssymb,epsfig}
\usepackage{latexsym}
\usepackage{amsmath}
\usepackage{graphicx}

\RequirePackage{color}

\begin{document}

\title{Generalized Chaplygin gas model:  Cosmological consequences and statefinder diagnosis}
\author{M. Malekjani, A. Khodam-Mohammadi  and
 N. Nazari-pooya} \affil{Department of Physics, Faculty of Science,
Bu-Ali Sina University, Hamedan 65178, Iran.\\}

\altaffiltext{}{malekjani@basu.ac.ir.}
\altaffiltext{}{khodam@basu.ac.ir.}
\altaffiltext{}{nazarip@basu.ac.ir.}

\begin{abstract}
The generalized Chaplygin gas (GCG) model in spatially flat universe
is investigated. The cosmological consequences led by GCG model
including the evolution of EoS parameter, deceleration parameter and
dimensionless Hubble parameter are calculated. We show that the GCG
model behaves as a general quintessence model. The GCG model can
also represent the pressureless CDM model at the early time and
cosmological constant model at the late time. The dependency of
transition from decelerated expansion to accelerated expansion on
the parameters of model is investigated. The statefinder parameters
$r$ and $s$ in this model are derived and the evolutionary
trajectories in $s-r$ plane are plotted. Finally, based on current
observational data, we plot the evolutionary trajectories in $s-r$
and $q-r$ planes for best fit values of the parameters of GCG model.
It has been shown that although, there are similarities between GCG
model and other forms of chaplygin gas in statefinder plane, but the
distance of this model from the $\Lambda$CDM fixed point in $s-r$
diagram is shorter compare with standard chaplygin gas model.
\end{abstract}

\keywords{Dark energy, Generalized Chaplygin gas model, Statefinder
diagnosis}
 \maketitle

\section{Introduction\label{intro}}
Since 1998, the type Ia supernova (SNe Ia) observations have shown
that the universe has undergone to the accelerating expansion phase
\citep{Riess98,Perlmuter99}. This fact has also been supported by
many additional observations, including the anisotropy measurements
of Cosmic Microeave Background (CMB) from Wilkinson Microwave
Anisotropy Probe (WMAP) \citep{Spergel03,Spergel06}, the data of
Large Scale Structure of universe (LSS) from Sloan Digital Sky
Survey (SDSS) \citep{Tegmark1,Tegmark2} and X-ray experiments
\citep{alen}. The current accelerating expansion of universe
indicates that in addition to the existence of dark matter, which is
required to explain the galactic dynamics and the formation of
structures \citep{bos81}, the universe is dominated by an exotic
energy component with negative pressure, dubbed the dark energy. In
an another word, in the framework of standard cosmology, the dark
energy (DE) scenario is a theoretical solution to explain the
accelerating expansion of the universe. The combined analysis of
cosmological observations suggest that the universe consists of
about 70\% dark energy, 30\% dust matter (cold dark matter plus
baryons), and negligible radiation. The cosmological constant, whose
equation of state is independent of cosmic time, is a simple
solution of DE problem. However, it suffers from two well known
problems, namely, the fine-tuning and the cosmic coincidence
problems \citep{copel}. In addition to cosmological constant, many
kinds of dynamical DE models, whose equation of state is no longer a
constant but slightly evolves with time, have been suggested to
interpret the cosmic acceleration. The quintessence  \citep{wet88},
phantom field \citep{cald02}, quintom \citep{eliza04}, Chaplygin gas
models \citep{kame01}, K-essence \citep{chi00}, tachyon field
\citep{sen02}, holographic \citep{coh99} and agegraphic \citep{Cai1}
DE models are the examples of dynamical DE model. It is emphasized
that the predictions of cosmological constant model is still fitted
to the observation \citep{LCDM1}. Therefore, a suggested dynamical
DE model should not be faraway from cosmological constant. Besides
the DE models, modified gravity theories such as scalar tensor
cosmology \citep{Boisseau}, braneworld models \citep{Dvali} have
been
suggested to solve the accelerated expansion of universe.\\
The Chaplygin gas is one of the candidate of DE models to explain
the accelerated expansion of the universe. The striking features of
Chaplygin gas DE is that it can be assumed as a possible unification
of dark matter and DE. The Chaplygin gas plays a dual role at
different epoch of the history of the universe: it can be as a
dust-like matter in the early time, and as a cosmological constant
at the late time. This model from the field theory points of view
has been investigated in \citep{Bil}. The Chaplygin gas emerges as
an effective fluid associated with D-branes \citep{Bor} and can also
be obtained from the Born-Infeld action \citep{Ben}. The simplest
form of Chaplygin gas model called standard Chaplygin gas (SCG)
which has been used to explain the accelerated expansion of universe
\citep{gori}. Although the SCG model can interpret the accelerated
expansion of universe, but it can not explain the astrophysical
problems such as the structure formation and cosmological
perturbation power spectrum \citep{sanvik}. Subsequently, the SCG is
extended into the generalized Chaplygin gas (GCG) which could
construct viable cosmological models. Same as SCG model, the GCG
model can obtain the accelerated expansion of the universe
\citep{setar}.

The quantities $H=\dot{a}/a$ and $q=-\ddot{a}/aH^2$, namely, the
Hubble parameter and the deceleration parameter, are the geometrical
parameters to describe the expansion history of universe, where $a$
is the scale factor and dot denotes the derivative with respect to
time. It is obvious that $\dot{a}>0 (H>0)$ indicates the expansion
of universe and $\ddot{a}>0 (q<0)$ means that the universe is
undergoing an accelerated expansion. The various DE models give the
same value of $q_0$ at present time, therefore, the hubble parameter
$H$ (first time derivative of scale factor) and the deceleration
parameter $q$ (second time derivative of scale factor) can not
discriminate the various DE models. For this aim, we need the higher
order of time derivative of scale factor. Using the third order time
derivative, Sahni, et al. \citep{Sahni} and Alam et al.\citep{alam}
introduced the statefinder pair \{r,s\} in order to remove the
degeneracy of $H_0$ and $q_0$ of  different DE models. The
statefinder pair \{r,s\} is defined as
\begin{equation}
r=\frac{\dddot{a}}{aH^{3}},~~~~~~~s=\frac{r-1}{3(q-1/2)},
\label{rspair}
\end{equation}
It is clear that the statefinder is a geometrical diagnostic,
because it depends only on the scale factor. The role of statefinder
pair is to distinguish the behaviors of cosmological evolution of
dark energy models with the same values of $H_{0}$ and $q_{0}$ at
the present time.
 Up to now, the statefinder diagnostic tool has been used to study
the various dark energy models. The various DE models have different
evolutionary trajectories in \{r,s\} plane, therefore the
statefinder is a good tool to discriminate DE models. For example,
the well-known $\Lambda$CDM model corresponds to a fixed point
$\{r=1,s=0\}$ in \{r,s\} plane \citep{Sahni}. Also, the quintessence
DE model \citep{Sahni,r10}, the interacting quintessence models
\citep{r12,r13}, the holographic dark energy models
\citep{r14,r15}, the holographic dark energy model in non-flat universe \citep%
{r16}, the phantom model \citep{r18}, the tachyon \citep{r22}, the
agegraphic DE model with and without interaction in flat and
non-flat universe \citep{wei,malek} and the interacting new
agegraphic DE model in flat and non-flat universe
\citep{zhang,khod10}, are analyzed through the statefinder
diagnostic tool. The SCG model is analyzed in terms of statefinder
with or without dust component \citep{mos}. In 2003, the generalized
cosmic chaplygin gas (GCCG) model is introduced by Gonzalez-Diaz
\citep{gon}. The interesting features of GCCG model is that it can
be stable and free from un physical behaviors even when the vacuum
fluid satisfies the phantom energy condition \citep{gon}.
 Chakraborty, et al. (2007) have
performed the statefinder analysis for GCCG model and for particular
choice of interaction parameter, they have shown the role of
statefinder parameters for the evolution of the universe
\citep{chak}. Zhang, et al., (2006) proposed a new model to describe
the unification of dark matter and DE, namely the new
 generalized chaplygin gas (NGCG) model and calculated the cosmological consequences and statefinder analysis for this
 model. The advantage of NGCG model is that it can represent the
 other forms of dark energy models such as quintessence-like and
 phantom-like dark energy \citep{zhang3}. \\
In this paper, first we study the cosmological consequences of
 GCG model, and then examine it by means of statefinder diagnostic
 tool. Here, we also study the dependency of the cosmological
 quantities and statefinder diagnostic on the parameters of GCG
 model. The paper is organized as follows: In section 2, we introduce the
GCG model and derive the statefinder parameters \{r,s\} for this
model. In section 3, the numerical results are presented. We
conclude in section 4.

\section{GCG model \label%
{theory}}
The equation of state of GCG is given by
\begin{equation}\label{state1}
p=-\frac{A}{\rho^{\alpha}}
\end{equation}
where $A>0$ and $\alpha\geq0$ are the parameters of the model
\citep{bento02}. In the case of $\alpha=1$, the GCG model is reduced
to SCG model. Also, for $\alpha=0$ it can be reduced to standard
$\Lambda$CDM model. In the framework of Friedmann- Robertson- Walker
(FRW) cosmology, using Eq. (\ref{state1}) and the conservation
equation $d(\rho a^3)=-pd(a^3)$, the energy density of GCG is
written as
\begin{equation}
\rho _{GCG}=\rho _{0GCG}[A_s+(1-A_s)a^{-3(1+\alpha )}]^{\frac
1{1+\alpha }},\label{e2}
\end{equation}
where $a$ is the scale factor, $A_s=A/\rho_{0GCG}^{1+\alpha}$ and
$\rho _{0GCG}$ is the present value of energy density. Using Eq.
(\ref{state1}) and (\ref{e2}), the equation of state (EoS) parameter
of GCG model can be obtained as
\begin{equation}\label{gcg_w}
w_{GCG}=-\frac{A_s a^{3(1+\alpha)}}{1-A_s+A_s a^{3(1+\alpha)}}
\end{equation}
From Eq. (\ref{gcg_w}), it is clear to show that at the early time
($a\rightarrow 0$), the EoS parameter tends to zero
($w_{GCG}\rightarrow 0$) and at the late time ($a\rightarrow
\infty$): $w_{GCG}\rightarrow -1$, which are equal to the EoS
parameter of matter and cosmological constant, respectively.
Therefore, the EoS parameter of GCG model is constrained to the
interval $-1\leq w_{\Lambda}\leq0$ and can be viewed as a general
quintessence DE model. Moreover, Eq. (\ref{gcg_w}) shows that at the
early time the GCG fluid can be interpreted as a CDM
($w_{GCG}\rightarrow 0$), and at the late time it mimics the
$\Lambda$CDM model ($w_{GCG}\rightarrow -1$). We can also see that
for $A_s>1$ the EoS
parameter $w_{GCG}<-1$ and the GCG model can cross the phantom divide.\\
 Since the dark matter and dark energy are unified by GCG model, therefore it can
be de-composited into two components of dark matter and dark energy,
as $\rho_{GCG}=\rho_{de}+\rho_{dm}$. Also, by assuming the
pressureless CDM, we have $p_{GCG}=p_{de}$. Considering the evolving
density of CDM as
\begin{equation}
\rho_{dm}=\rho_{0dm}a^{-3},
\end{equation}
it is obvious that the energy density of DE in GCG model can be
derived as
\begin{eqnarray}
&\rho_{de}&=\rho_{GCG}-\rho_{dm}=\\
&&\rho_{0GCG}[A_s+(1-A_s)a^{-3(1+\alpha )}]^{\frac 1{1+\alpha
}}-\rho _{0dm}a^{-3}.\nonumber \label{e4}
\end{eqnarray}
 By assuming that the universe
is filled by GCG component (DE+CDM) and baryonic matter component,
the total energy density is $\rho_t=\rho_{GCG}+\rho_b$. In the case
of flat universe, the friedmann equation for GCG model is written as
\begin{eqnarray}\label{hub1}
&H^2&=\frac{8\pi G}{3}\rho_t\\
 &&=\frac{8\pi G}{3}\left(\rho
_{0GCG}[A_s+(1-A_s)a^{-3(1+\alpha )}]^{\frac 1{1+\alpha
}}+\rho_{0b}a^{-3}\right)\nonumber
\end{eqnarray}
 where
$\rho_{0b}$ is the present density of baryonic matter. Substituting
the dimensionless parameters
\begin{eqnarray}
\rho_{0GCG}=\frac{3H_0^2}{8\pi G}\Omega_{GCG} &&\nonumber\\
\rho_{0b}=\frac{3H_0^2}{8\pi G}\Omega_{b} \\
\Omega_{GCG}+\Omega_b=1, \nonumber
\end{eqnarray}
in Eq. (\ref{hub1}), the Hubble parameter is expressed as
\begin{eqnarray}
&H^2&=H_0^2E^2(a)=\\
&&H_0^2\Big((1-\Omega_{b})[A_s+(1-A_s)a^{-3(1+\alpha)}]^{\frac{1}{1+\alpha}}+\Omega_{b}a^{-3}\Big),\nonumber
\end{eqnarray}
 where $E(a)$, the the normalized Hubble parameter, is
defined as
\begin{equation}\label{E1}
E(a)=\left(
(1-\Omega_{b})[A_s+(1-A_s)a^{-3(1+\alpha)}]^{\frac{1}{1+\alpha}}+\Omega_{b}a^{-3}\right)^{1/2}
\end{equation}
Recently, Xu and Lu \citep{xu}, by applying the Markov Chain Monte
Carlo approach on the latest observational data, have constrained
the GCG model. The observational data that have been used are: the
constitution dataset \citep{397Constitution} including 397 type
supernova Ia (SNIa), the observational Hubble data (OHD)
\citep{OHD}, the cluster X-ray gas mass fraction
\citep{ref:07060033}, the measurement results of baryon acoustic
oscillation (BAO) from Sloan Digital Sky Survey (SDSS) \citep{SDSS}
and Two Degree Field Galaxy Redshift Survey (2dFGRS)
\citep{ref:Percival2}, and the cosmic microwave background (CMB)
data from five-year WMAP \citep{5yWMAP}. They obtained that in the
flat universe, the best fit values of the GCG model parameters
($A_s$, $\alpha$) and the cosmological parameters ($\Omega_bh^2$,
$H_0$) with their confidence level are:
$A_{s}=0.76^{+0.029}_{-0.039}$ ($1\sigma$) $^{+0.034}_{-0.046}$
$(2\sigma)$, $\alpha=0.033^{+0.066}_{-0.071}$ ($1\sigma$)
$^{+0.096}_{-0.087}$ $(2\sigma)$,
$\Omega_bh^2=0.0233^{+0.0023}_{-0.0016}$ ($1\sigma$)
$^{+0.0029}_{-0.0020}$ $(2\sigma)$ and $H_0=69.97^{+2.87}_{-2.78}$
($1\sigma$) $^{+3.48}_{-3.08}$ $(2\sigma)$, with minimum chi-square
$\chi^2_{min}=519.342$. At following, we derive the deceleration
parameter $q$ and the statefinder pair \{r,s\} for GCG model.\\
The deceleration parameter $q$, which denotes the expansion phase of
the universe, is given by
\begin{equation}
q=-\frac{\dot{H}}{H^2}-1
\end{equation}
Re-witting $q$ in terms of $E$, we have
\begin{equation}\label{q1}
q(x)=-\frac{1}{E}\frac{dE}{d\ln{a}}-1
\end{equation}
Substituting  $E$ from Eq. (\ref{E1}) in (\ref{q1}), we obtain the
deceleration parameter $q$ for GCG model as

\begin{eqnarray}\label{q_5}
&q&=
\Big[3(1-\Omega_b)\Big(A_s+(1-A_s)a^{-3(1+\alpha)}\Big)^{-\frac{\alpha}{1+\alpha}}
\nonumber\\&&\times (1-A_s)a^{-3(1+\alpha)}+3\Omega_ba^{-3}\Big]\Big/\nonumber\\
 &&\Big[2(1-\Omega_b)\Big(A_s+(1-A_s)a^{-3(1+\alpha)}\Big)^\frac{1}{1+\alpha}\nonumber\\
 &&+2\Omega_ba^{-3}\Big]-1.
\end{eqnarray}

Eq. (\ref{q_5}) explicitly shows the dependence of deceleration
parameter $q$ on the GCG model parameter $A_s$ and $\alpha$. Now, we
derive the statefinder pair \{r,s\} for GCG model. Using the
definition of statfinder parameters in Eq. (\ref{rspair}), we have
\begin{equation}
r=\frac{\dddot{a}}{aH^3}=\frac{\ddot{H}}{H^3}-3q-2.
\end{equation}
Similar to $q$, the parameter $r$ can be re-written in terms of $E$
as
\begin{equation}\label{r1}
r=\frac{1}{E}\frac{d^2E}{d(\ln{a})^2}+\frac{1}{E^2}(\frac{dE}{d\ln{a}})^2+\frac{3}{E}\frac{dE}{d\ln{a}}+1,
\end{equation}

also the parameter $s$ can be obtained as
\begin{equation}\label{s1}
s=-\frac{\frac{1}{E}\frac{d^2E}{d(\ln{a})^2}+\frac{1}{E^2}(\frac{dE}{d\ln{a}})^2+\frac{3}{E}\frac{dE}{d\ln{a}}}{\frac{3}{E}\frac{dE}{d\ln{a}}+\frac{9}{2}}
\end{equation}
Substituting $E$ from Eq. (\ref{E1}) in (\ref{r1}) and (\ref{s1}),
we obtain $r$ and $s$ for GCG model as

\begin{eqnarray}\label{r2}
&r&= 1+\Big[9(1-\Omega_b)(1-A_s)\times \\
&&\Big(A_s+(1-A_s)a^{-3(1+\alpha)}\Big)^{-\frac{1+2\alpha}{1+\alpha}}
A_s \alpha a^{-3(1+\alpha)}\Big]\Big/ \nonumber\\
&&\Big[2(1-\Omega_b)\Big(A_s+(1-A_s)a^{-3(1+\alpha)}\Big)^\frac{1}{1+\alpha}+2\Omega_ba^{-3}\Big]
\nonumber,
\end{eqnarray}

\begin{equation}\label{s2}
s=-\frac{(1-A_s)\alpha
a^{-3(1+\alpha)}}{A_s+(1-A_s)a^{-3(1+\alpha)}}
\end{equation}
For $\alpha=0$ or $A_s=1$, we obtain $\{r=1, s=0\}$ which refers to
the statefinder pair of spatially flat $\Lambda$CDM model. Departure
of a given dark energy model from the fixed point $\{r=1,s=0\}$ is a
criterion for evaluating of this model from spatially flat
$\Lambda$CDM model \citep{Sahni}. The importance of the statefinder
diagnostic is that the current values of the parameters $s$ and $r$
can be extracted from the observational data of SNAP (Super Nova
Acceleration Probe) type experiments. Therefore, the statefinder
diagnostic combined with the future SNAP observation can help us to
discriminate between different dark energy models.

\section{Numerical results\label{NR}}
In this section we discuss the cosmological consequences led by GCG
model. For this aim, the evolution of EoS parameter of GCG model,
$w_{GCG}$, the deceleration parameter $q$ and cosmological evolution
of dimensionless hubble parameter ,$E$,  are studied. Then we study
the GCG model by means of statefinder diagnostic point of view.
\subsection{EoS parameter}
 Solving Eq. (\ref{gcg_w}), the evolution of $w_{GCG}$
as a function of scale factor for different model parameters $A_s$
and $\alpha$ is shown in Fig. (1). In upper panel, by fixing
$A_s=0.76$ based on observational constrain, we vary the parameter
$\alpha$ as $1, 0.1, 0.01$. For $a<1$, increasing the parameter
$\alpha$ leads to a larger value of $w_{GCG}$. While, at $a>1$,
$w_{GCG}$ is smaller for larger value of $\alpha$. Here, we see the
dual role of GCG model at different epoch of the history of
universe: It can be assumed as a pressureless dark matter ($w=0.0$)
at the early time and a cosmological constant with $w=-1$ at the
late time. In lower panel, by fixing $\alpha=0.033$ based on
observational constrain, we plot $w_{GCG}$ for different
illustrative values of $A_s$. For $A_s=0.0$, we see that
$w_{GCG}=0.0$, which refers to the pressureless CDM model. In the
case of $A_s=1.0$, we have $w_{GCG}=-1$, which denotes the
cosmological constant. For $0<A_s<1$, we can see $-1<w_{GCG}<0$,
denoting the time varying equation of state of GCG model and
representing the general quintessence behavior of this model. Also,
increasing the parameter $A_s$ leads to smaller value of $w_{GCG}$.
\subsection{Deceleration parameter}
Here we calculate the evolution of deceleration parameter, $q$ for a
universe dominated by GCG model and investigate the dependency of
$q$ on the parameters of model. In Fig.(2), by solving
Eq.(\ref{q_5}), the evolution of $q$ as a function of scale factor
for different illustrative values of model parameters $\alpha$ and
$A_s$ is calculated. Here we adopt the observational value of
$\Omega_b h^2$ as $0.0233$. First we fix the parameter $A_s$ as
$0.76$ and vary the parameter $\alpha$ as $1.0$, $0.1$ and $0.01$
(upper panel). Then, by fixing $\alpha=0.033$, we choose the
illustrative values for $A_s$ as $0.0$, $0.1$, $0.2$, $0.3$ and
$1.0$ (lower panel). In upper panel, one can see that by increasing
the parameter $\alpha$ $q$ becomes larger for $a<1$ and smaller for
$a>1$. The transition from decelerated expansion ($q>0$) to
accelerated expansion ($q<0$) takes place earlier for smaller value
of $\alpha$. From this figure we see that $q=1/2$ at the early time
(CDM dominated universe) and $q$ tends to $-1$ at the late time
($\Lambda$CDM dominated universe). This fact can also be seen in the
lower panel for any value of $A_s$ in the interval $0<A_s<1$. For
$A_s=0.0$, we have the constant deceleration parameter $q=1/2$,
corresponding to the CDM dominated universe. For the values in the
interval $0<A_s<1$, $q$ starts from $1/2$ at the early time and
leads to $-1$ at the late time. Furthermore, $q$ becomes larger for
lower values of $A_s$. Transition from decelerated expansion ($q>0$)
to accelerated expansion ($q<0$) occurs sooner for higher values of
$A_s$.
\subsection{Hubble parameter}
 At following, we study the Hubble parameter which evaluates the
 expansion rate of the universe for GCG cosmology. Using Eq. (\ref{E1}), we plot
the cosmological evolution of $E(a)$ in Fig. (3). First, we fix the
coefficient $A_s=0.76$ and vary the model parameter $\alpha$ as
$1.0$, $0.1$ and $0.01$ (see upper panel). In this case, we see that
the larger value the parameter $\alpha$ is taken, the bigger value
the Hubble expansion rate $E(a)$ gets. In lower panel, by fixing
$\alpha$, we vary the coefficient $A_s$ as $0.0$, $0.1$, $0.2$,
$0.3$ and $1.0$. The case of $A_s=1.0$ represents the standard
$\Lambda$CDM model and $A_s=0.0$ denotes the CDM model. Here, one
can see that for larger value of $A_s$, the Hubble expansion rate
$E(a)$ becomes smaller at $a<1$ and larger at $a>1$. Therefore, from
the above analysis, we find that both parameters $A_s$ and $\alpha$
can impact the cosmic expansion history in GCG model.
\subsection{Statefinder parameters}
 The statefinder pair \{r,s\}
for GCG model is given by Eqs. (\ref{r2}) and (\ref{s2}),
respectively. In statefinder plane, the horizontal axis is defined
by the parameter $s$ and vertical axis by the parameter $r$. In this
diagram, the standard $\Lambda$CDM model corresponds to a fixed
point $\{r=1,s=0\}$. At the early time, $a\rightarrow 0$, the
statefinder pair $\{r,s\}$ defined in Eqs. (\ref{r2}) and (\ref{s2})
are reduced as follows:
\begin{equation}
r=\frac{9}{2}\frac{(1-\Omega_b)(1-A_s)^{-\frac{1+2\alpha}{1+\alpha}}A_s\alpha
a^{3(\alpha+1)}}{(1-\Omega_b)(1-A_s)^{\frac{1}{1+\alpha}}}+1,~~~s=-\alpha,
\end{equation}
therefore we see that at the early time, $a\rightarrow 0$, the
statefinder pair \{r,s\} for GCG model are: $\{s=-\alpha, r=1\}$.
From Eqs. (\ref{r2}) and (\ref{s2}), we can also obtain the
statefinder pair \{r,s\} at the late time, $a\rightarrow \infty$, as
$\{r=1, s=0.0\}$. Hence, the GCG model mimics the cosmological
constant at the late time. In Fig.(4), we show the evolutionary
trajectories of GCG model in statefinder plane. In upper panel, by
fixing $A_s=0.76$, we choose the illustrative values $0.1$, $0.2$
and $0.3$ for $\alpha$. While the universe expands, the trajectories
of the statefinder start from the fixed point $\{s=-\alpha, r=1\}$
at the early time. The parameter $r$ starts to increase and then
decares, while the parameter $s$ increases from the initial value
$s=-\alpha$ at the early time to $s=0.0$ at the late time. Here, we
can easily see that the statefinder trajectory is dependent on the
parameter $\alpha$ of GCG model. Different values of $\alpha$ give
the different evolutionary trajectories in $\{s, r\}$ plane. The
colore points on the curves represent the today's values of
statefinder parameters ($s_0,r_0$) and the star symbol indicates the
standard $\Lambda$CDM model. The distance to $\Lambda$CDM fixed
point becomes shorter for smaller value of $\alpha$. We can also see
that for smaller values of $\alpha$, the parameter $s$ increases and
the parameter $r$ decreases. In lower panel, by fixing
$\alpha=0.033$, we plot the evolutionary trajectories in $s-r$
diagram for different values of the parameter $A_s$. From Eqs. Eqs.
(\ref{r2}) and (\ref{s2}), we have $\{ r=1,s=-\alpha\}$ for
$A_s=0.0$. Therefore, for $A_s=0.0$ we have a fixed point
\{$s=-\alpha, r=1$\} in $s-r$ plane. Also form Eqs. (\ref{r2}) and
(\ref{s2}) one can see that \{$s=0, r=1$\} for $A_s=1.0$. Hence, in
the case of $A_s=1.0$, the statefinder parameters \{$s,r$\} are
coincide to the $\Lambda$CDM fixed point in $s-r$ plane. For
different values of $A_s$ in the interval $0<A_s<1$, we have the
different evolutionary trajectories in $s-r$ plane (see the right
panel). Therefore, the parameter $A_s$ affects the evolutionary
trajectories in $s-r$ plane. The distance to $\Lambda$CDM fixed
point \{$s=0.0,r=1.0$\} becomes shorter for larger values of $A_s$.
The color points represent the today's values of the parameters
($s_0,r_0$) for different values of $A_s$. Here, we see that the
parameter $s_0$ increases for larger values of $A_s$ and the
parameter $r_0$ is largest for $r=0.5$. In Fig. (5), we plot the
evolutionary trajectory in $s-r$ plane (upper panel) and $q-r$ plane
(lower panel) for the best fit observational values: $A_s=0.76$ and
$\alpha=0.033$. In upper panel the evolutionary trajectory starts
from ($s=-0.033,r=1$) at the past time, reaches to the
($s=-0.033,r=1$) at the present time (circle point) and ended at
($s=0,r=1$) at the future. This behavior of GCG model in statefider
plane is similar to NGCG model at the early time \citep{zhang3},
where they found that the universe starts from the initial value
($r=1$, $s=-\alpha$) in $s-r$ plane. Gorini, et al, \citep{mos}
calculated the trajectory of SCG in $s-r$ plane and showed that the
universe in SCG model starts form ($s=-1,r=1$) reaches to
($s_0=-0.3,r=1.9$) at the present time and finally mimics the
$\Lambda$CDM model at the late time. Therefore the distance of
($s_0,r_0$)in GCG model constrained by the above observational value
from the standard fixed point $\Lambda$CDM ($s=0,r=1$) is shorter
compare with SCG model. As a similarity, we see that for both model
the universe mimics the $\Lambda$CDM model at the late time.
Moreover, the behavior of the trajectory is $s-r$ plane is similar
for both model, where by expanding the universe, $r$ increases to a
maximum value then decreases to $r=1$ at the late time and the
parameter $s$ increases forever. In lower panel the evolutionary
trajectory in $q-r$ plane starts from ($q=0.5,r=1$) at the past
(note that ($q=0.5,r=1$) corresponds to the CDM dominated universe),
reaches to ($q=-0.575,r=1.26$) at the present time and ended at
($q=-0.965,r=1$) at the future.
\section{Conclusion}
Summarizing this work, we investigated the generalized Chaplygin gas
(GCG) model in spatially flat universe. Here we studied the
cosmological consequences of GCG model by calculating the evolution
of EoS parameter $w_{GCG}$, deceleration parameter $q$ and
cosmological evolution $E(a)$ of GCG model. In the GCG cosmology,
the universe starts from the CDM-dominated phase at the early time
to the DE-dominated universe at the late time.  The GCG fluid as a
general quintessence dark energy model can be viewed as a
pressureless matter fluid ($w_{GCG}=0.0$) at the early time and as a
cosmological constant ($w_{GCG}=-1$) at the late time. We also
obtained the deceleration parameter $q$ in GCG model and studied the
evolutionary treatment of $q$ as a function of scale factor in this
model. In GCG model, the parameter $q$ starts from the initial value
$1/2$ at the early time (CDM-dominated universe) and converges to
$-1$ at the late time ($\Lambda$-dominated universe). Furthermore,
we exhibit the cosmological evolution of $E(a)$ (normalized Hubble
parameter, $E(a)=H(a)/H_0$). For GCG model, both the parameters
$A_s$ and $\alpha$ affect the cosmological evolution. Finally, we
performed the statefinder diagnostic tool on the GCG model. Since
many cosmological models have been proposed to interpret the
accelerated expansion of universe, the statefinder diagnostic tool
with the parameters $r$ and $s$ which are constructed by higher
order derivative of the scale factor is needed to discriminate
between them. Moreover, the present values of $r$ and $s$ can be
viewed as a discriminator for testing a given dark energy model if
it can be extracted from observational data in a model-independent
way. Here we derived the statefinder parameters $r$ and $s$ for GCG
model and studied the evolutionary trajectories of this model in
$s-r$ plane. The dependence of the evolutionary trajectories and the
today's value of \{s,r\} on the model parameters $A_s$ and $\alpha$
has been investigated. The lower value of $\alpha$ and higher value
of $A_s$ result the shorter distance from standard $\Lambda$CDM
model in $s-r$ diagram. Eventually, we plotted the evolutionary
trajectory of GCG model in $s-r$ and $q-r$ plane based on current
observational data and found that the distance of GCG model from the
standard $\Lambda$CDM model in $s-r$ plane is shorter compare with
SCG model. However, both SCG and GCG models have a similar
trajectories in $s-r$ diagram.Furthermore, the behavior of GCG model
in statefider plane is similar to NGCG model at the early time
\citep{zhang3}, where the universe expands from the initial value
($r=1$, $s=-\alpha$) in $s-r$ plane.
\newpage
\begin{center}
\begin{figure}[!htb]
\includegraphics[width=8cm]{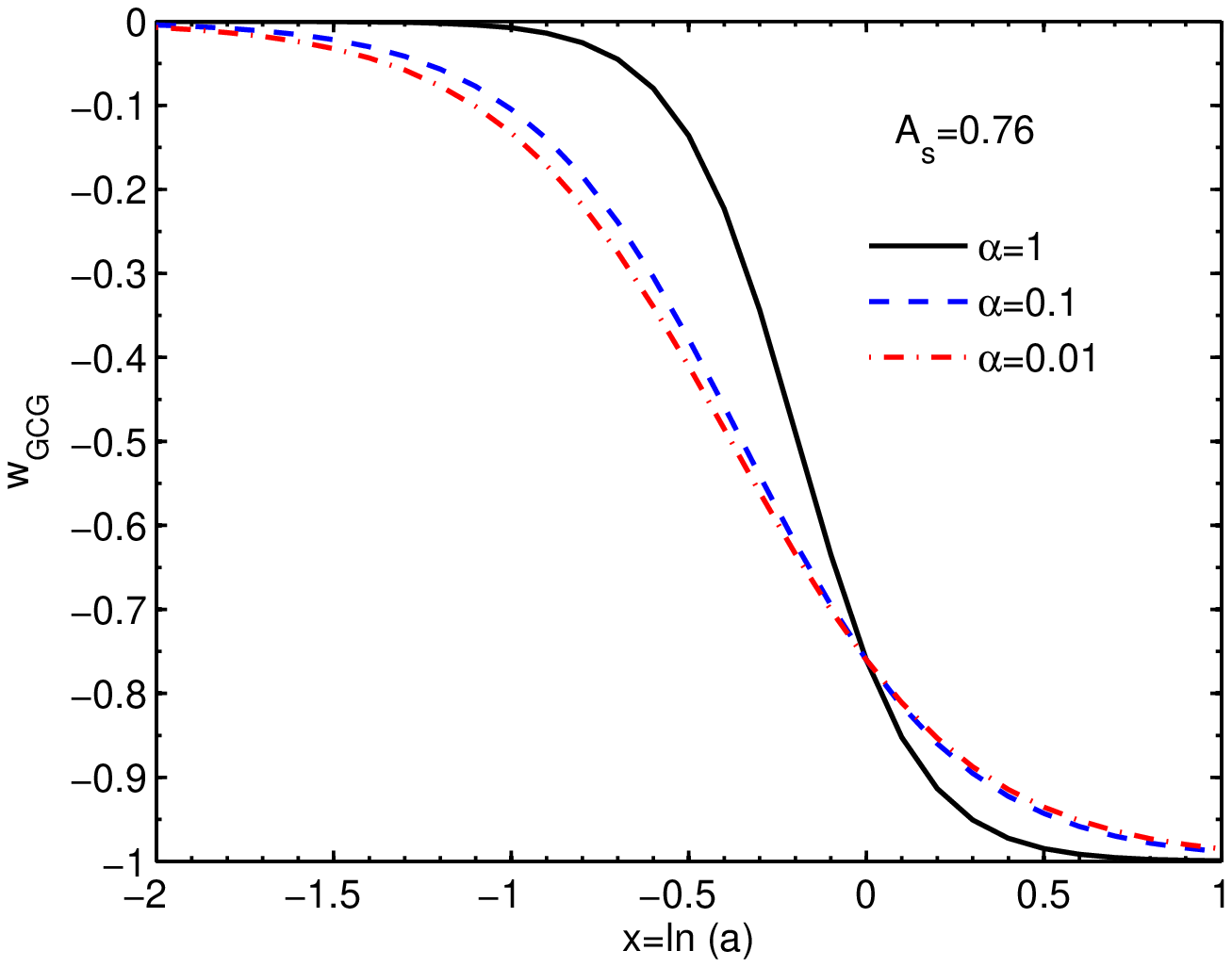} \includegraphics[width=8cm]{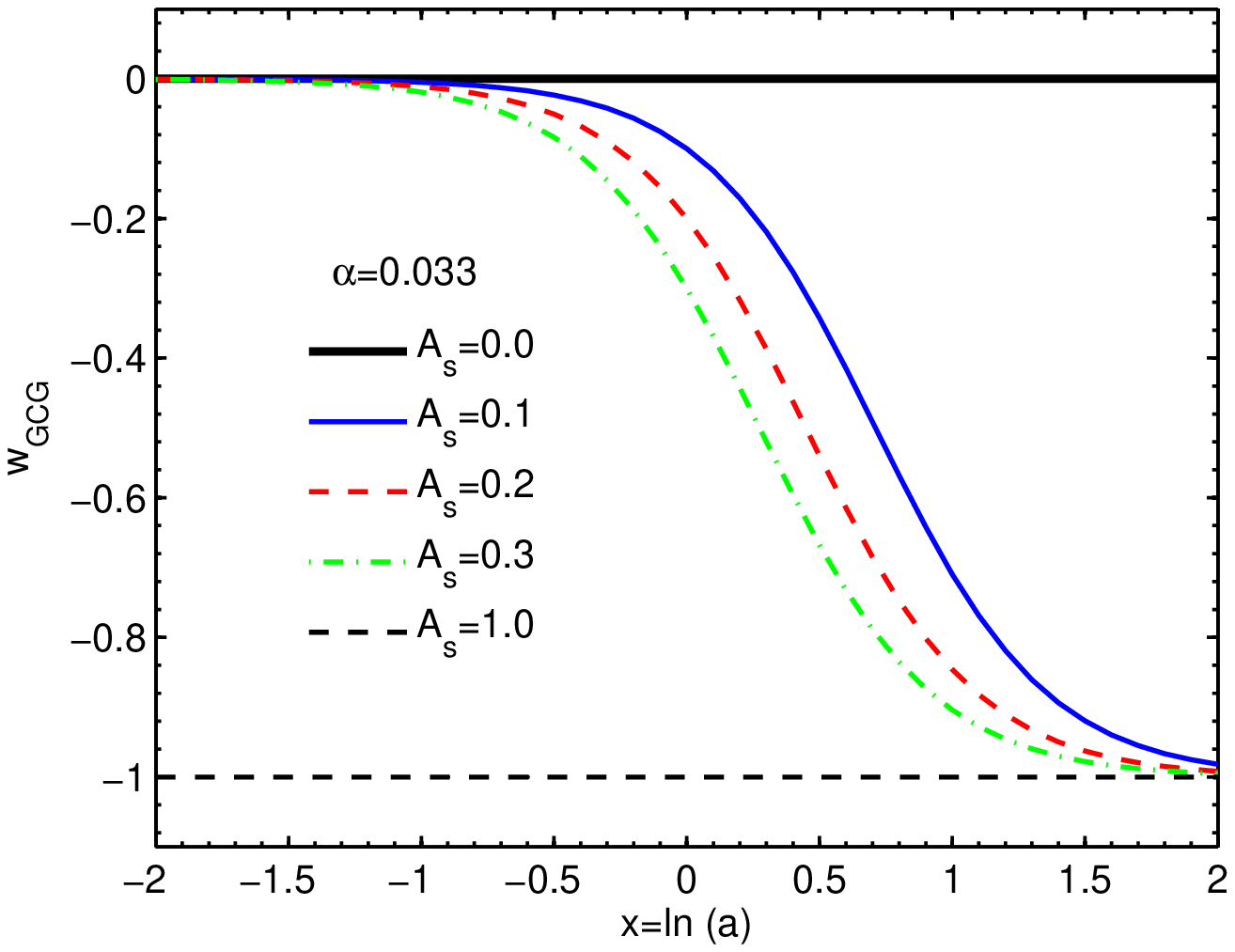} %
\caption{The evolution of EoS parameter of GCG model, $w_{GCG}$,
versus $x=\ln{a}$ for different illustrative values of parameters
$A_s$ and $\alpha$. In upper panel, fixing $A_s$ by best fit value:
$A_s=0.76$, we vary $\alpha$ as $1, 0.1 , 0.01$ corresponding to
black solid line, blue dashed line and red dotted-dashed line,
respectively. The case of $\alpha=1.0$ exhibits the standard
Chaplygin gas (SCG) model. In lower panel, fixing $\alpha$ by best
fit value: $\alpha=0.033$, $A_s$ is varied as $0$ (black thick solid
line), $0.1$ (blue solid line), $0.2$ (red dashed line), $0.3$
(green dotted-dashed line) and $1.0$ (black dashed line). The cases
of $A_s=0.0$ and $A_s=1.0$ exhibit the EoS parameter of pressureless
matter and cosmological constant fluids, respectively.}
\end{figure}
\end{center}

\newpage
\begin{center}
\begin{figure}[!htb]
\includegraphics[width=8cm]{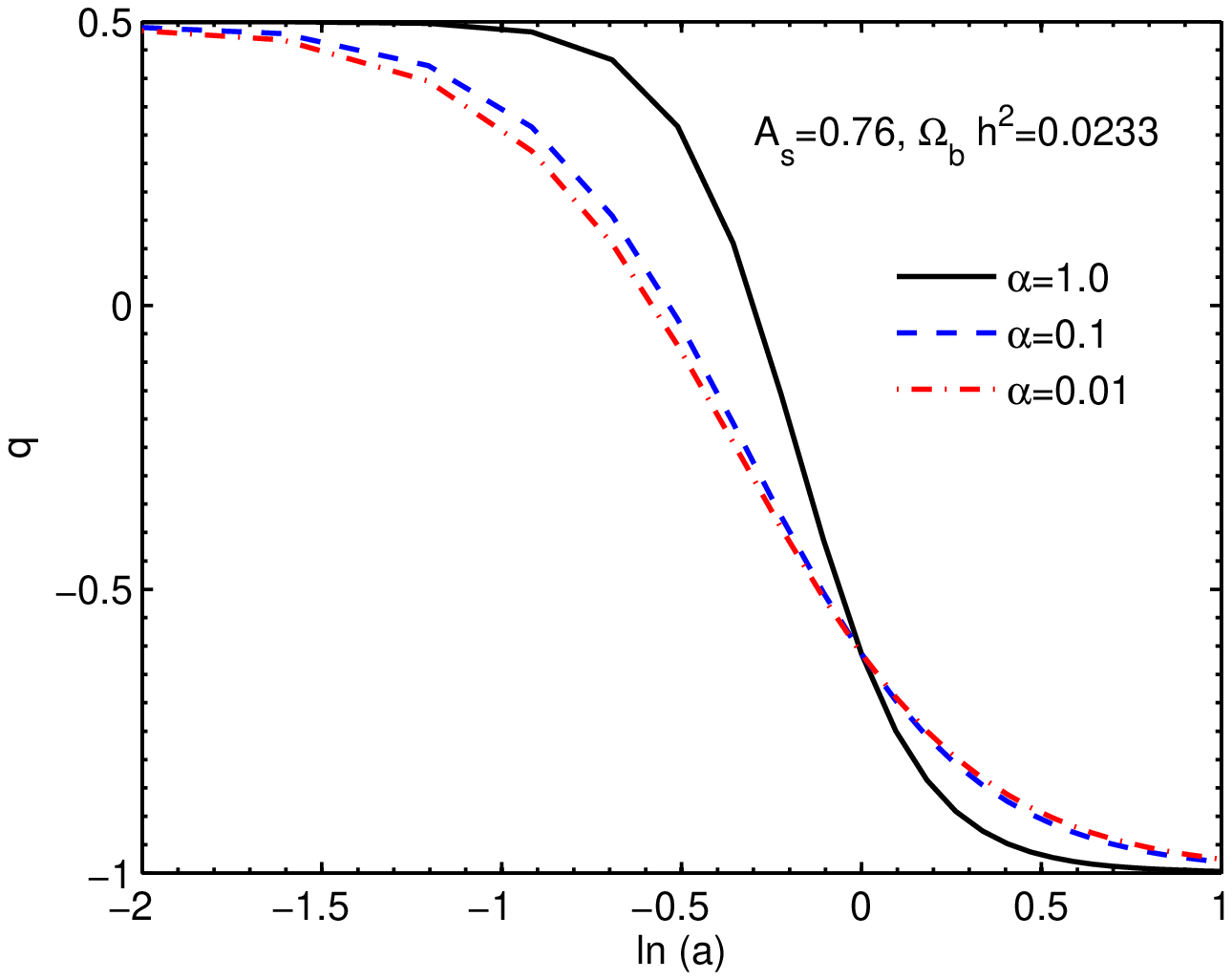} \includegraphics[width=8cm]{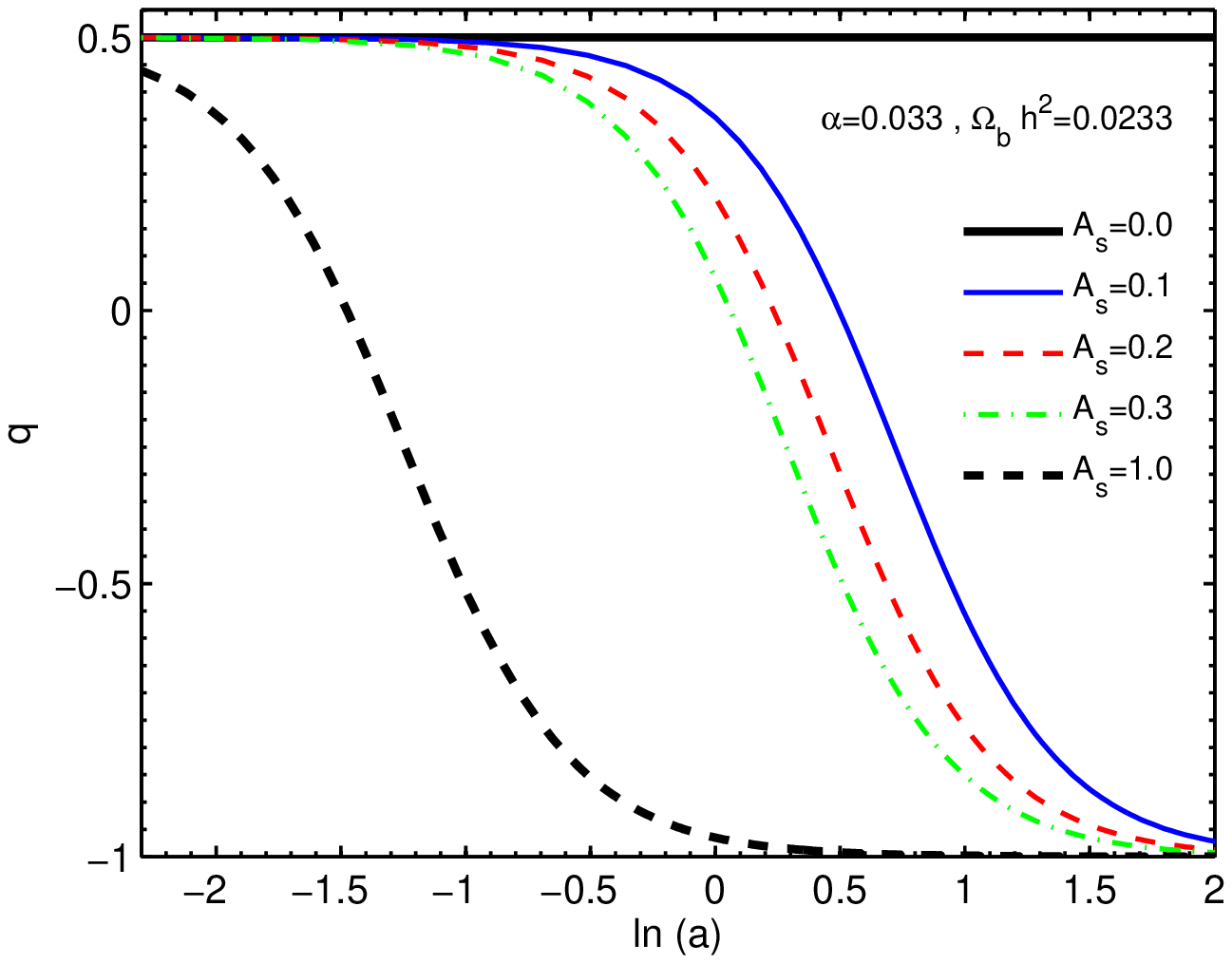} %
\caption{The evolution of deceleration parameter $q$ in GCG model
versus $x=\ln{a}$ for different illustrative values of model
parameters $A_s$ and $\alpha$. In upper panel, by fixing $A_s=0.76$,
we vary $\alpha$ as $0.2, 0.1 , 0.01$ corresponding to black solid
line, blue dashed line and red dotted-dashed line, respectively. The
case of $\alpha=1.0$ represents the SCG model.
 In
lower panel, by fixing $\alpha=0.033$, $A_s$ is varied as $0$ (black
thick solid line), $0.1$ (blue solid line), $0.2$ (red dashed line),
$0.3$ (green dotted-dashed line) and $1.$ (black thick dashed line).
The cases of $A_s=0.0$ and $A_s=1.0$ represent the evolution of $q$
in CDM-dominated and $\Lambda$-dominated universe, respectively.}
\end{figure}
\end{center}
\newpage
\begin{center}
\begin{figure}[!htb]
\includegraphics[width=8cm]{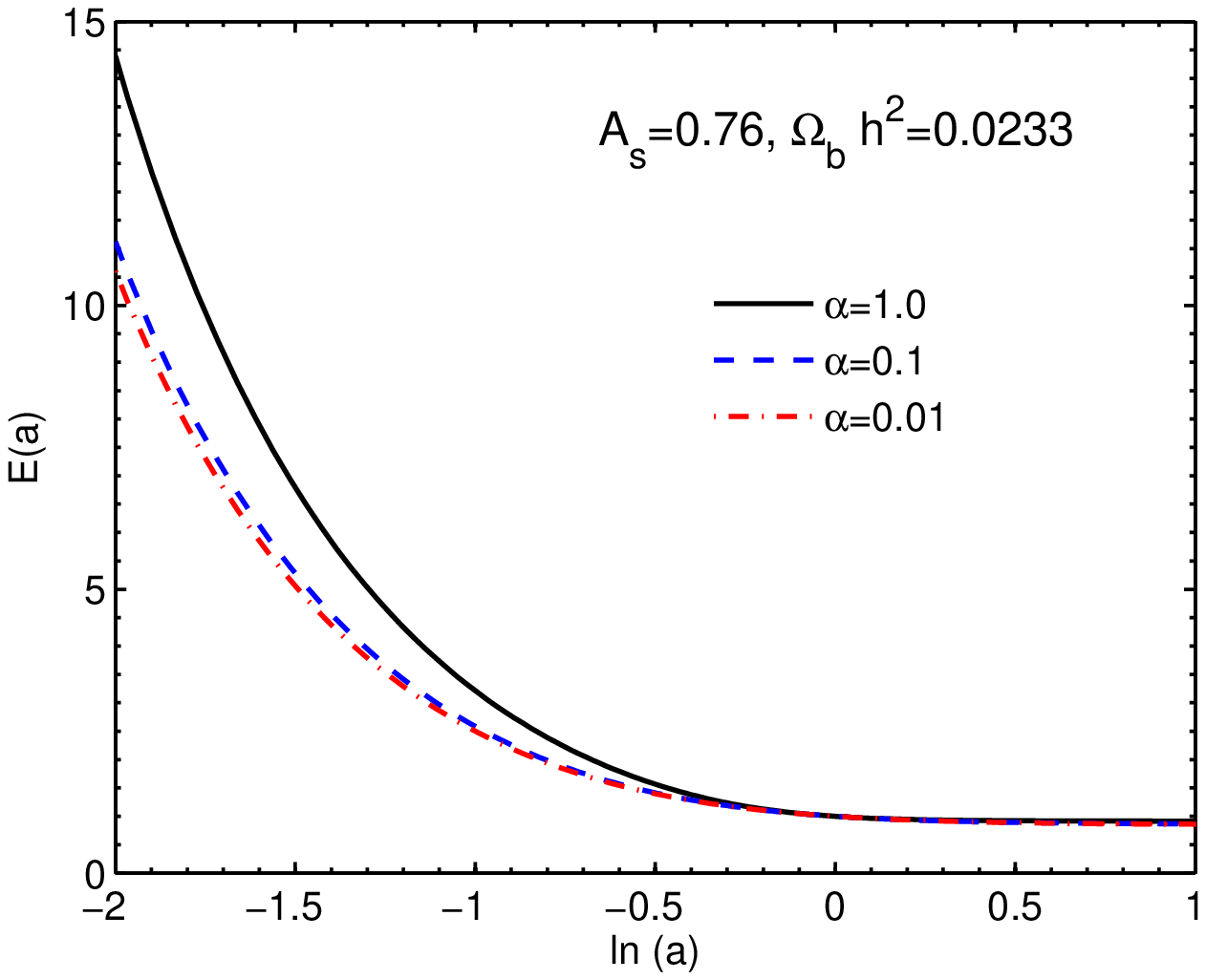} \includegraphics[width=8cm]{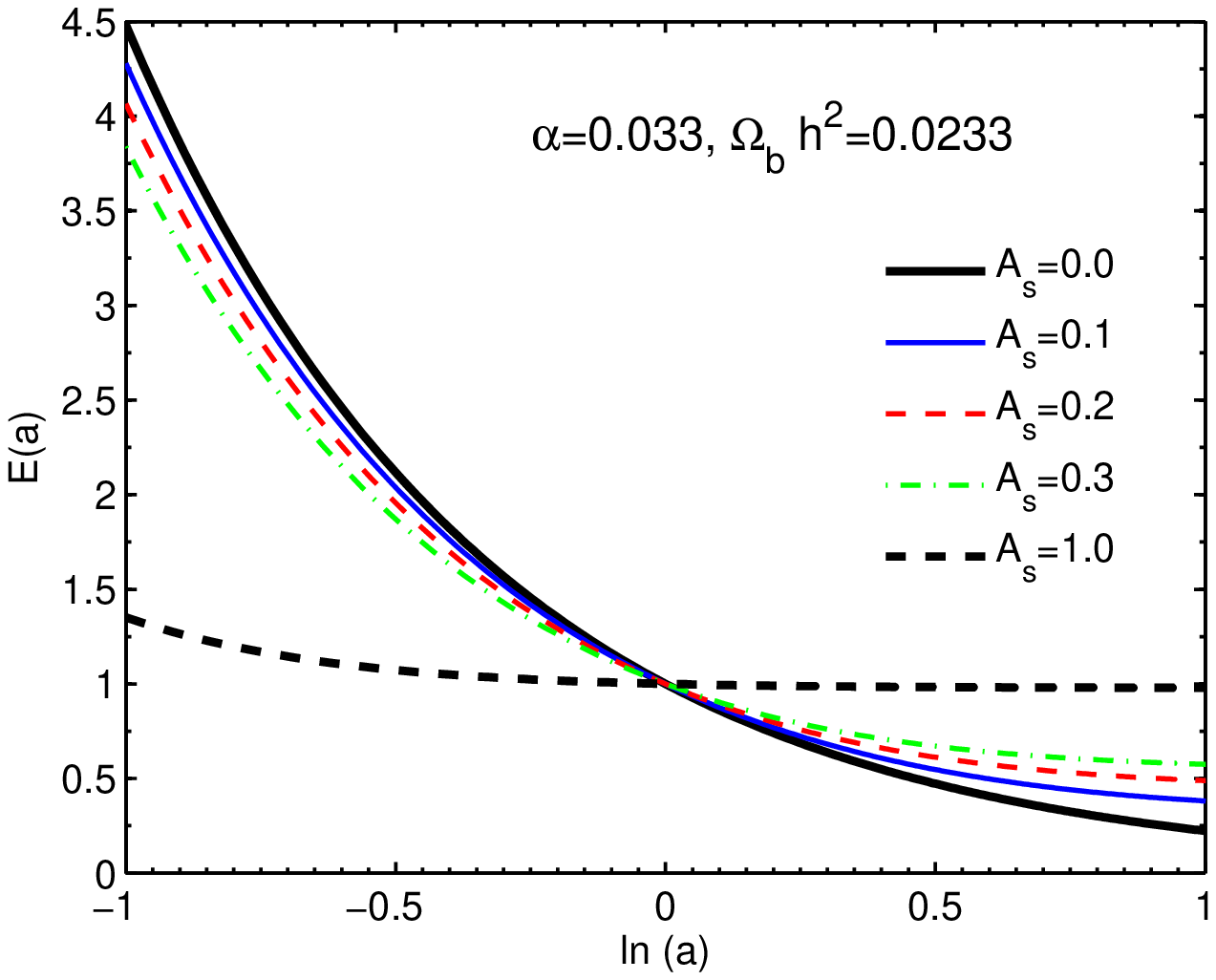} %
\caption{ Cosmological evolution of normalized Hubble parameter as a
function of logarithmic scale factor $x=\ln{a}$ in GCG model. In
uper panel, we choose the observational best fit values: $A_s=0.76$
and $\Omega_b h^2=0.0233$ and vary the parameter $\alpha$ as $1.0$,
$0.1$ and $0.01$ corresponding to black solid, blue dashed and red
dotted-dashed lines, respectively. The case of $\alpha=1.0$ exhibit
the SCG model. In lower panel, by fixing $\alpha=0.033$, $A_s$ is
varied as $0.0$, $0.1$, $0.2$, $0.3$ and $1.0$ corresponding to
black thick solid line, blue solid line, red dashed line, green
dotted-dashed line, black thick dashed line, respectively.}
\end{figure}
\end{center}
\newpage

\begin{center}
\begin{figure}[!htb]
\includegraphics[width=8cm]{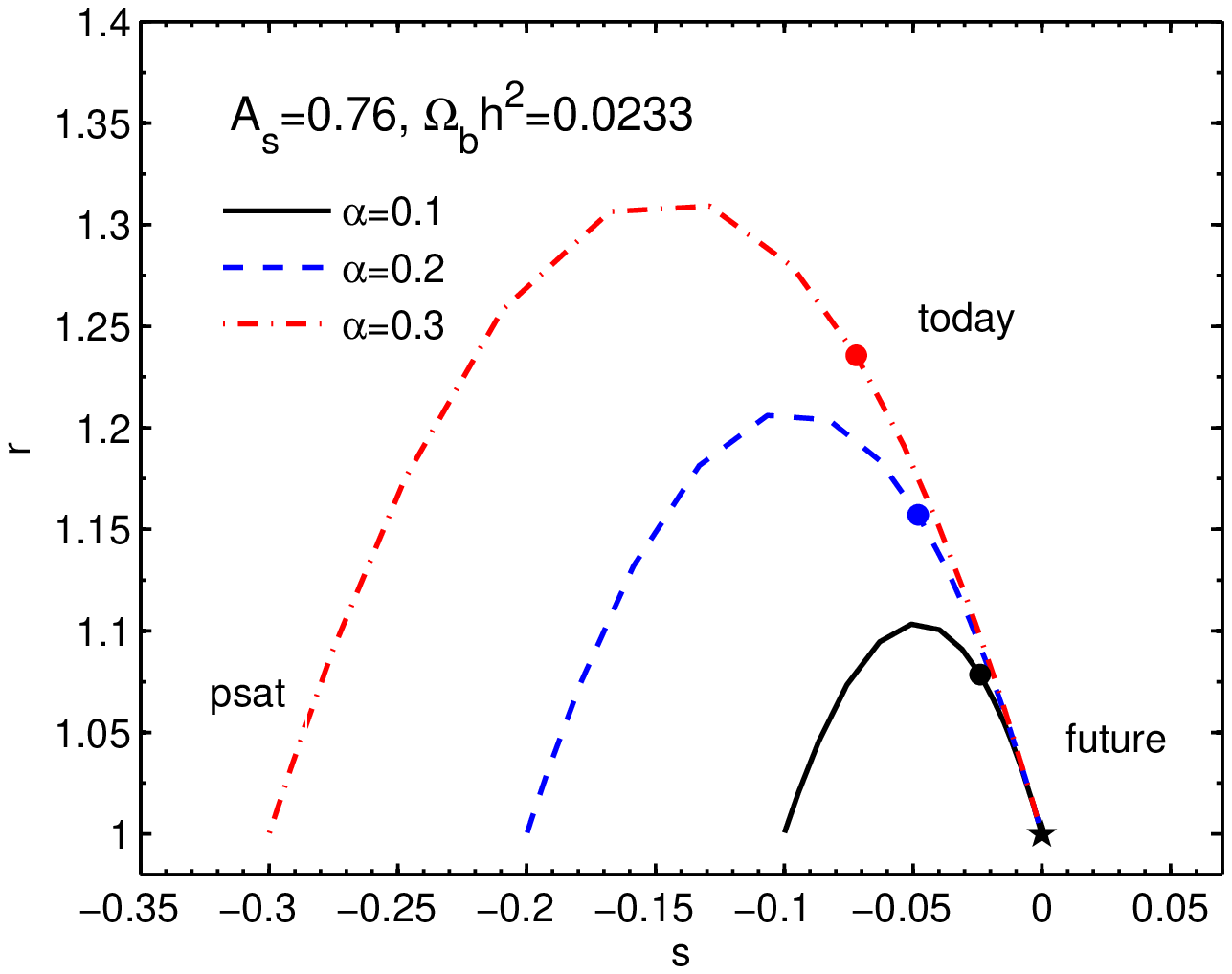} \includegraphics[width=8cm]{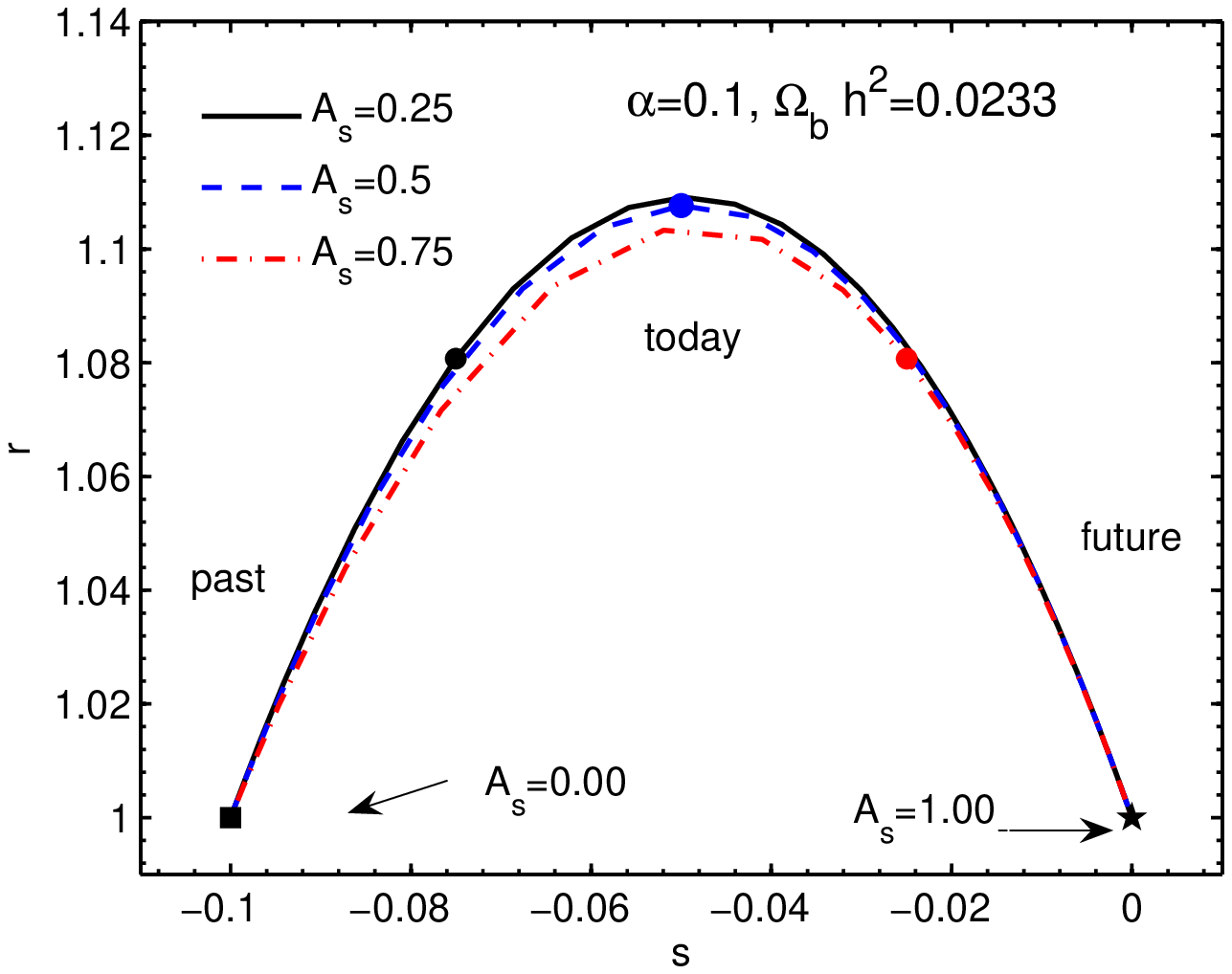} %
\caption{An illustrative example for the statefinder diagnostic of
GCG model. In upper panel, the evolutionary trajectories in $s-r$
plane are plotted, by fixing $A_s=0.76$ and varying $\alpha$ as
$0.1$, $0.2$ and $0.3$ corresponding to black solid line, blue
dashed line and red dotted dashed line, respectively. The circle
point on the curves show the today's value of statefinder parameters
($s_0,r_0$). The star symbol indicates the location of standard
$\Lambda$CDM model in $s-r$ plane:$\{s=0,r=1\}$. In lower panel, the
evolutionary trajectories are plotted for different illustrative
values of $A_s$, by fixing $\alpha=0.1$. The case of $A_s=0.0$ is
related to square symbol located at $\{s=-0.1,r=1.0\}$. Note that
$A_s=0.0$ represents the CDM model. The case of $A_s=1.0$ is
exhibited by star symbol at: $\{s=0,r=1.0\}$ which is related to
$\Lambda$CDM model. The evolutionary trajectories of illustrative
cases $A_s=0.25$, $A_s=0.50$ and $A_s=0.75$ have been shown by black
solid line, blue dashed line and red dotted-dashed line,
respectively. Circle point on the curves denotes the today's value
($s_0,r_0$) in $s-r$ plane.}
\end{figure}
\end{center}
\newpage
\begin{center}
\begin{figure}[!htb]
\includegraphics[width=8cm]{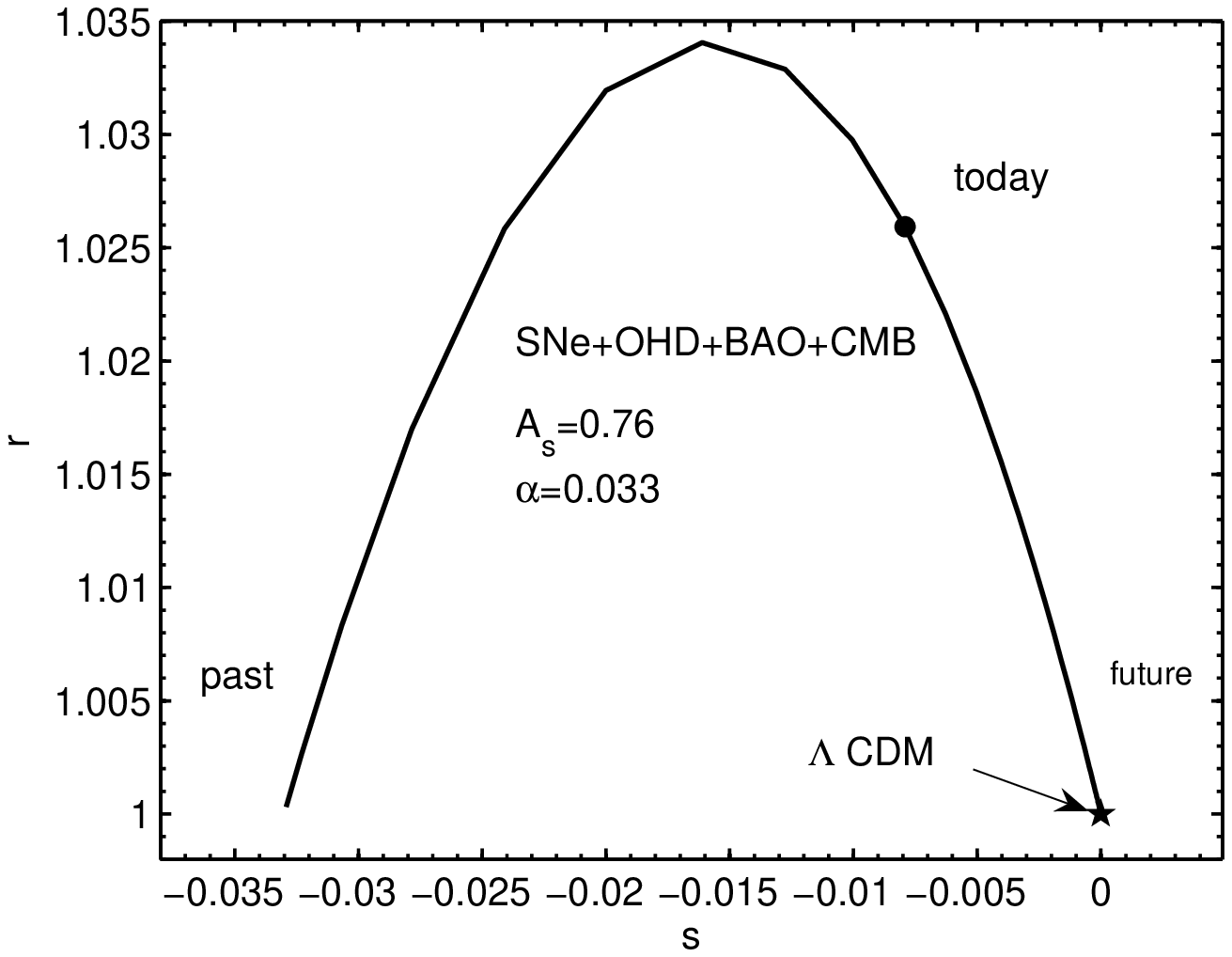} \includegraphics[width=8cm]{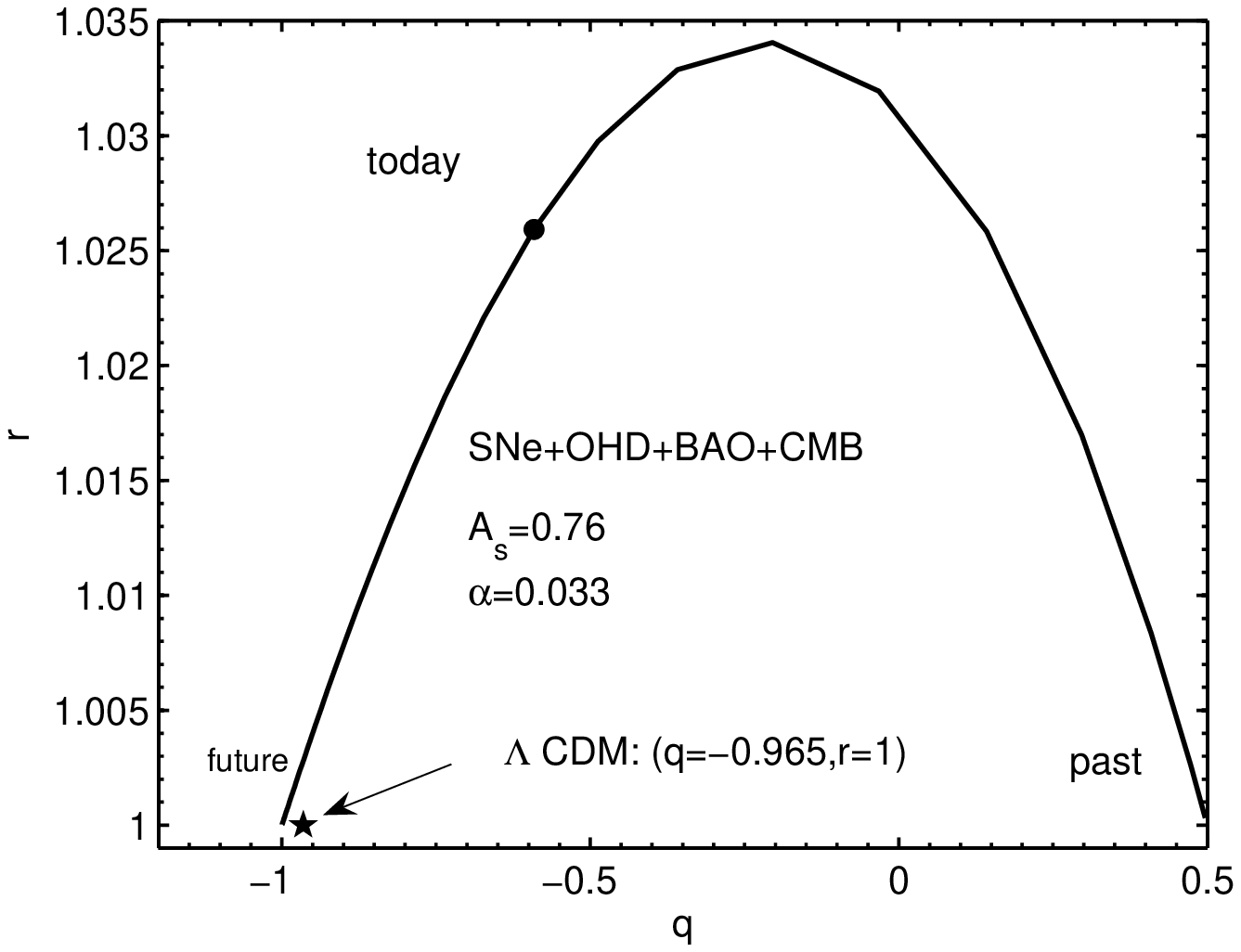} %
\caption{The statefinder diagrams $r(s)$ (upper panel) and $r(q)$
(lower panel) for GCG model. The evolutionary trajectories are
plotted in the light of best fit result of SNe+OHD+BAO+CMB,
$\alpha=0.033$ and $A_s=0.76$. The circlepoints on the curves show
the today's value ($s_0,r_0$), upper panel, and ($q_0,r_0$), lower
panel. For comparison, the standard $\Lambda$CDM model has been
shown by star symbol in these diagrams.}
\end{figure}
\end{center}

\end{document}